\documentclass[reprint,aps,prl,amsmath,amssymb,superscriptaddress,showpacs,floatfix]{revtex4-1}
\usepackage{graphicx}
\usepackage{bm}
\usepackage{dcolumn}
\usepackage{amsfonts}
\usepackage{hyperref}
\usepackage[usenames]{color}
\DeclareMathOperator{\sech}{sech}
\newcommand{\scalefig}{0.393}
\begin{document}

\title{A unified first-principles study of Gilbert damping, spin-flip diffusion and resistivity in transition metal alloys}
\author{Anton A. Starikov}

\author{Paul J. Kelly}
\affiliation{Faculty of Science and Technology and MESA$^+$ Institute for Nanotechnology, University of Twente, P.O. Box 217, 7500 AE Enschede, The Netherlands}

\author{Arne Brataas}
\affiliation{Department of Physics, Norwegian University of Science and Technology, N-7491 Trondheim, Norway}

\author{Yaroslav Tserkovnyak}
\affiliation{Department of Physics and Astronomy, University of California, Los Angeles, California 90095, USA}

\author{Gerrit E. W. Bauer}
\affiliation{Kavli Institute of NanoScience, Delft University of Technology, Lorentzweg 1, 2628 CJ Delft, The Netherlands}

% \date{\today ~version 6c}
\date{\today}

\begin{abstract}
Using a formulation of first-principles scattering theory that includes disorder and spin-orbit coupling on an equal footing, we calculate the resistivity $\rho$, spin flip diffusion length $l_{\rm sf}$ and the Gilbert damping parameter $\alpha$ for Ni$_{1-x}$Fe$_x$ substitutional alloys as a function of $x$. For the technologically important Ni$_{80}$Fe$_{20}$ alloy, permalloy, we calculate values of 
$\rho = 3.5 \pm 0.15$~$ \mu$Ohm-cm, $l_{\rm sf}=5.5 \pm 0.3$~nm, and $\alpha= 0.0046 \pm 0.0001$ compared to experimental low-temperature values in the range 
$4.2-4.8$~$ \mu$Ohm-cm for $\rho$, $5.0-6.0$~nm for $l_{\rm sf}$, and $0.004-0.013$ for $\alpha$ indicating that the theoretical formalism captures the most important contributions to these parameters.
\end{abstract}

%
% Gilmore:prl07 PACS numbers: 75.40.Gb, 76.60.Es 
% Brataas:prl08 PACS numbers: 75.40.Gb, 72.25.Mk, 76.60.Es 
% Fedorov:prb08 PACS numbers: 71.70.Ej, 72.25.Rb, 76.30.-v, 85.75.-d 
%
% 71.70.Ej Spin-orbit coupling, Zeeman and Stark splitting, Jahn-Teller effect 
%
% 72.15.Gd Galvanomagnetic and other magnetotransport effects
%
% 72.25.-b Spin polarized transport (for spin polarized transport devices, see 85.75.-d)
%
% 72.25.Ba Spin polarized transport in metals
% 
% 72.25.Mk Spin transport through interfaces 
%
% 72.25.Rb Spin relaxation and scattering
% 
% 73.40.Jn Metal-to-metal contacts
% 
% 75.40.Gb Dynamic properties (dynamic susceptibility, spin waves, spin diffusion, dynamic scaling, etc.)
% 
% 75.47.De Giant magnetoresistance
% 
% 75.47.−m Magnetotransport phenomena; materials for magnetotransport
% 
% 75.70.Cn Magnetic properties of interfaces (multilayers, superlattices, heterostructures)
% 
% 75.70.Ak Magnetic properties of monolayers and thin films
%
% 75.75.+a Magnetic properties of nanostructures
% 
% 75.76.+j Spin transport effects
% 
% 76.30.-v Electron paramagnetic resonance and relaxation
%
% 76.50.+g Ferromagnetic, antiferromagnetic, and ferrimagnetic resonances; spin-wave resonance
% 
% 76.60.Es Relaxation effects
%
% 85.75.-d Magnetoelectronics; spintronics: devices exploiting spin polarized transport or integrated magnetic fields

\maketitle

\paragraph{\color{red} Introduction.}The drive to increase the density and speed of magnetic forms of data storage has focussed attention on how magnetization changes in response to external fields and currents, on shorter length- and time-scales \cite{[See the collection of articles ]UMS}. The dynamics of a magnetization ${\bf M}$ in an effective magnetic field ${\bf H}_{\rm eff}$ is commonly described using the phenomenological Landau-Lifshitz-Gilbert equation 
\begin{equation}
  \frac{d{\bf M}}{dt} =  - \gamma {\bf M} \times {\bf H}_{\rm eff} + {\bf M} \times 
  \left[ \frac{\tilde G({\bf M})}  {\gamma M_s^2} \frac{d{\bf M}}{dt} \right] \:, \label{eqn:LLG}
\end{equation}
where $M_s=|{\bf M}|$ is the saturation magnetization, ${\tilde G({\bf M})}$ is the Gilbert damping parameter (that is in general a tensor) and the gyromagnetic ratio $\gamma=g \mu_B/\hbar$ is expressed in terms of the Bohr magneton $\mu_B$ and the Land\'{e} $g$ factor, which is approximately 2 for itinerant ferromagnets. The {\it time} decay of a magnetization precession is frequently expressed in terms of the dimensionless parameter $\alpha$ given by the diagonal element of $ {\tilde G} / \gamma M_s $ for an isotropic medium. If a non-equilibrium magnetization is generated in a disordered metal (for example by injecting a current through an interface), its {\it spatial} decay is described by the diffusion equation
\begin{equation}
\frac{\partial^2 \Delta \mu}{\partial z^2} =  \frac{\Delta \mu} {l^2_{\rm sf}} \label{eqn:VF}
\end{equation}
in terms of the spin accumulation $\Delta \mu$, the difference between the spin-dependent electrochemical potentials $\mu_s$ for up and down spins, and the spin-flip diffusion length $l_{\rm sf}$ \cite{vanSon:prl87,*vanSon:prl88,Valet:prb93}. In spite of the great importance of $\alpha$ and $l_{\rm sf}$, our understanding of the factors that contribute to their numerical values is at best sketchy. For clean ferromagnetic metals \cite{Gilmore:prl07,*Gilmore:jap08,*Kambersky:prb07} and ordered alloys \cite{Liu:apl09} however, recent progress has been made in calculating the Gilbert damping using the Torque Correlation Model (TCM) \cite{Kambersky:cjp76} and the relaxation time approximation in the framework of the Boltzmann equation. Estimating the relaxation time for particular materials and scattering mechanisms is in general a non-trivial task and application of the TCM to non-periodic systems entails many additional complications and has not yet been demonstrated. Thus, the theoretical study of Gilbert damping or spin-flip scattering in disordered alloys and their calculation for particular materials with intrinsic disorder remain open questions.

\paragraph{\color{red} Method.}In this paper we calculate the resistivity $\rho$, spin-flip diffusion length $l_{\rm sf}$ and Gilbert damping parameter $\alpha$ for substitutional Ni$_{1-x}$Fe$_x$ alloys within a single first-principles framework. To do so, we have extended a scattering formalism \cite{Xia:prb01,*Xia:prb06} based upon the local spin density approximation (LSDA) of density functional theory (DFT) so that spin-orbit coupling (SOC) and chemical disorder are included on an equal footing. Relativistic effects are included by using the Pauli Hamiltonian. 

For a disordered region of ferromagnetic (F) alloy sandwiched between leads of non-magnetic (N) material, the scattering matrix $S$ relates incoming and outgoing states in terms of reflection  ($r$) and transmission matrices ($t$) at the Fermi energy. To calculate the scattering matrix, we use a ``wave-function matching'' (WFM) scheme \cite{Xia:prb01,*Xia:prb06} implemented with a minimal basis of tight-binding linearized muffin-tin orbitals (TB-LMTOs) \cite{Andersen:prb86,*Andersen:prb75}. Atomic-sphere-approximation (ASA) potentials \cite{Andersen:prb86,*Andersen:prb75} are calculated self-consistently using a surface Green's function (SGF) method also implemented \cite{Turek:97} with TB-LMTOs. Charge and spin densities for binary alloy $A$ and $B$ sites are calculated using the coherent potential approximation (CPA) \cite{Soven:pr67} generalized to layer structures \cite{Turek:97}. For the transmission matrix calculation, the resulting spherical potentials are assigned randomly to sites in large lateral supercells (SC) subject to maintenance of the appropriate concentration of the alloy \cite{Xia:prb01,*Xia:prb06}. Solving the transport problem using lateral supercells makes it possible to go beyond effective medium approximations such as the CPA. Because we are interested in the properties of bulk alloys, the leads can be chosen for convenience and we use Cu leads with a single scattering state for each value of crystal momentum, ${\bf k}_{\parallel}$. The alloy lattice constants are determined using Vegard's law and the lattice constants of the leads are made to match. Though NiFe is fcc only for the concentration range $0 \leq x \leq 0.6$, we use the fcc structure for all values of $x$.

For the self-consistent SGF calculations (without SOC), the two-dimensional (2D) Brillouin zone (BZ) corresponding to the $1\times1$ interface unit cell was sampled with a $120\times120$ grid. Transport calculations including spin-orbit coupling were performed with a $32\times32$ 2D BZ grid for a $5\times5$ lateral supercell, which is equivalent to a $160\times160$ grid in the $1\times1$ 2D BZ. The thickness of the ferromagnetic layer ranged from $3$ to  $200$ monolayers of fcc alloy; for the largest thicknesses, the scattering region contained more than 5000 atoms. For every thickness of ferromagnetic alloy, we averaged over a number of random disorder configurations; the sample to sample spread was small and typically only five configurations were necessary.

\begin{figure}[bth!]
\includegraphics[scale=\scalefig]{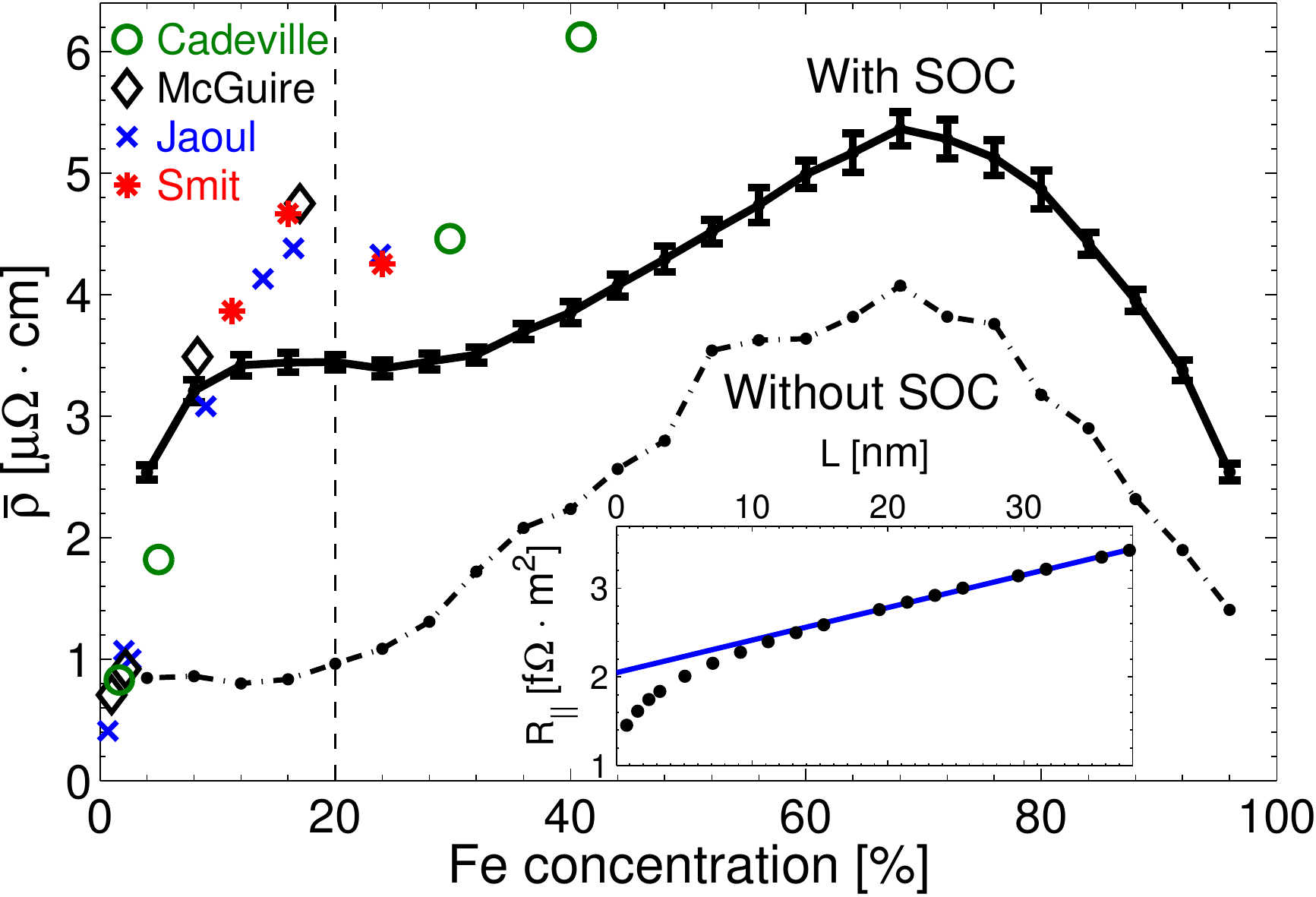}
\caption{
Calculated resistivity as a function of the concentration $x$ for fcc Ni$_{1-x}$Fe$_x$ binary alloys with (solid line) and without (dashed-dotted line) SOC. Low temperature experimental results are shown as symbols  \cite{Smit:phys51,*McGuire:ieeem75,*Jaoul:jmmm77,*Cadeville:jpf73}. The composition Ni$_{80}$Fe$_{20}$ is indicated by a vertical dashed line.
Inset: resistance of Cu$|$Ni$_{80}$Fe$_{20}|$Cu as a function of the thickness of the alloy layer. Dots indicate the calculated values averaged over five configurations while the solid line is a linear fit. 
}
\label{fig:Fig1}
\end{figure}

\paragraph{\color{red} Resistivity.}We calculate the electrical resistivity to illustrate our methodology. In the Landauer-B{\"ut}tiker formalism, the conductance can be expressed in terms of the transmission matrix $t$ as $ G = (e^2/h) \text{Tr} \left\{ tt^\dag \right\} $ \cite{Buttiker:prb85,Datta:95}.
The resistance of the complete system consisting of ideal leads sandwiching a layer 
of ferromagnetic alloy of thickness $L$ is 
$ R(L) = 1/G(L) =  1/G_{\rm Sh} + 2 R_{\rm if} + R_{\rm b}(L) $
where 
$G_{\rm Sh}= \left( 2 e^2/h \right) N $
is the Sharvin conductance of each lead with $N$ conductance channels per spin,
$R_{\rm if}$ is the interface resistance of a single N$|$F interface, and 
$R_{\rm b}(L)$ is the bulk resistance of a ferromagnetic layer of thickness $L$ \cite{Schep:prb97,Xia:prb06}. When the ferromagnetic slab is sufficiently thick, Ohmic behaviour is recovered whereby $R_{\rm b}(L)\approx \rho L$ as shown in the inset to Fig.~\ref{fig:Fig1} for permalloy (Py = Ni$_{80}$Fe$_{20}$) and the bulk resistivity $\rho$ can be extracted from the slope of $R(L)$ \cite{fn2}. For currents parallel and perpendicular to the magnetization direction, the resistivities are different and have to be calculated separately. The average resistivity is given by $\bar{\rho}=(\rho_{\parallel}+2\rho_{\perp})/3$, and the anisotropic magnetoresistance ratio (AMR) is $(\rho_{\parallel}-\rho_{\perp})/\bar{\rho}$.

For permalloy we find values of $\bar{\rho}= 3.5 \pm 0.15$~$\mu$Ohm-cm and AMR $=19 \pm 1 \%$, compared to experimental low-temperature values in the range $4.2-4.8$~$\mu$Ohm-cm for $\bar{\rho}$ and $18\%$ for AMR \cite{Smit:phys51}. The resistivity calculated as a function of $x$ is compared to low temperature literature values \cite{Smit:phys51,*McGuire:ieeem75,*Jaoul:jmmm77,*Cadeville:jpf73} in Fig.~\ref{fig:Fig1}. The plateau in the calculated values around the Py composition appears to be seen in the experiments by Smit and Jaoul {\em et al.} \cite{Smit:phys51,*Jaoul:jmmm77}. The overall agreement with previous calculations is good \cite{Banhart:epl95,*Banhart:prb97}. In spite of the smallness of the SOC, the resistivity of Py is underestimated by more than a factor of four when it is omitted, underlining its importance for understanding transport properties.

%%%%%%%%%%%%%%%%%%%%%%%%%%%%%%%%%%%%%%%%%%%%%%%%%%%%%%%%%%%%%%%%%%%%%%%%%%%%%%%%%%%%%%%%%%%%
%%%%%%%%%%%%%%%%%%%%%%%%%%%%%%%%%%%%%%%%%%%%%%%%%%%%%%%%%%%%%%%%%%%%%%%%%%%%%%%%%%%%%%%%%%%%
%%%%%%%%%%%%%%%%%%%%%%%%%%%%%%%%%%%%%%%%%%%%%%%%%%%%%%%%%%%%%%%%%%%%%%%%%%%%%%%%%%%%%%%%%%%%
%

Three sources of disorder which have not been taken into account here will increase the calculated values of $\rho$; short range potential fluctuations that go beyond the single site CPA, short range strain fluctuations reflecting the differing volumes of Fe and Ni and spin disorder. These will be the subject of a later study.

% subsection resistivity (end)

\paragraph{\color{red} Gilbert Damping.}
Recently, Brataas {\it et al.} showed that the energy loss due to Gilbert damping in an N$|$F$|$N scattering configuration can be expressed in terms of the scattering matrix $S$ \cite{Brataas:prl08}. Using the Landau-Lifshitz-Gilbert equation (\ref{eqn:LLG}), the energy lost by the ferromagnetic slab is,
\begin{equation}
  \frac{dE}{dt} = \frac{d}{dt} \left( {\bf H}_{\rm eff} \cdot {\bf M} \right) = {\bf H}_{\rm eff} \cdot \frac{d{\bf M}}{dt}= \frac{1}{\gamma^2} \frac{d{\bf m}}{dt} {\tilde G({\bf M})} \frac{d{\bf m}}{dt} 
\end{equation}
where ${\bf m}={\bf M}/M_s$ is the unit vector of the magnetization direction for the macrospin mode. By equating this energy loss to the energy flow into the leads \cite{Avron:prl01,*Moskalets:prb02a,*Moskalets:prb02b} associated with ``spin-pumping'' \cite{Tserkovnyak:prl02a,*Tserkovnyak:prb02b},
\begin{equation}
   I_E^{Pump} = \frac{\hbar}{4\pi} \text{Tr} \left\{ {\frac{dS}{dt} \frac{dS^\dag}{dt}} \right\} = \frac{\hbar}{4\pi} \text{Tr} \left\{ \frac{dS}{d{\bf m}} \frac{d{\bf m}}{dt} \frac{dS^\dag}{d{\bf m}} \frac{d{\bf m}}{dt} \right\}\:,
 \end{equation}
the elements of the tensor ${\tilde G}$ can be expressed as
\begin{equation}
  \tilde G_{i,j} ({\bf m}) = \frac{\gamma ^2 \hbar}{4\pi}{\mathop{\rm Re}\nolimits} \left\{ 
\text{Tr} \left[ \frac{\partial S}{\partial m_i}\frac{\partial S^\dag}{\partial m_j} \right] \right\} \:. 
\label{eqn:dampeq}
\end{equation}
Physically, energy is transferred slowly from the spin degrees of freedom to the electronic orbital degrees of freedom from where it is rapidly lost to the phonon degrees of freedom. Our calculations focus on the role of elastic scattering in the rate-limiting first step.

\begin{figure}[btp]
\includegraphics[scale=\scalefig]{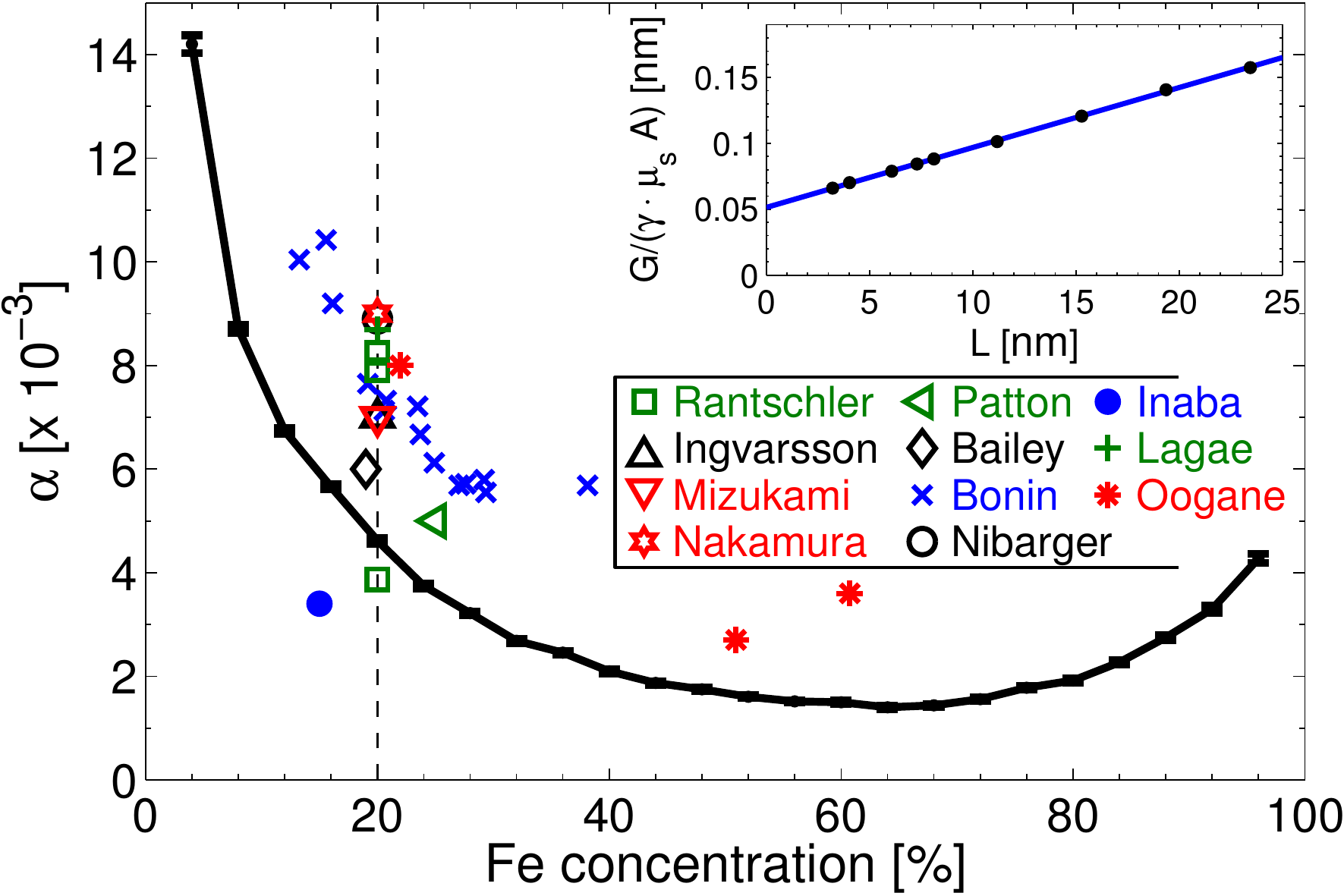}
\caption{Calculated zero temperature (solid line) and experimental room temperature (symbols) values of the Gilbert damping parameter as a function of the concentration $x$ for fcc Ni$_{1-x}$Fe$_x$ binary alloys \cite{Mizukami:jmmm01,*Mizukami:jjap01,Bailey:ieeem01,
Patton:jap75,*Ingvarsson:apl04,*Nakamura:jjap04,*Rantschler:ieeem05,*Bonin:jap05,
*Lagae:jmmm05,*Nibarger:apl03,*Inaba:ieeem06,*Oogane:jjap06}.  
Inset: total damping of Cu$|$Ni$_{80}$Fe$_{20}|$Cu as a function of the thickness of the alloy layer. Dots indicate the calculated values averaged over five configurations while the solid line is a linear fit. 
}
\label{fig:Fig2}
\end{figure}

Assuming that the Gilbert damping is isotropic for cubic substitutional alloys and allowing for the enhancement of the damping due to the F$|$N interfaces \cite{Tserkovnyak:prl02a,Zwierzycki:prb05,Mizukami:jmmm01,*Mizukami:jjap01}, the total damping in the system with a ferromagnetic slab of thickness $L$ can be written
$  {\tilde G}(L)={\tilde G}_{\rm if}+{\tilde G}_b(L) $
where we express the bulk damping in terms of the dimensionless Gilbert damping parameter 
$ {\tilde G_b}(L)=\alpha \gamma M_s(L) = \alpha \gamma \mu_s A L$,
where $\mu_s$ is the magnetization density and $A$ is the cross section. The results of calculations for Ni$_{80}$Fe$_{20}$ are shown in the inset to Fig.~\ref{fig:Fig2}, where the derivatives of the scattering matrix in \eqref{eqn:dampeq} were evaluated numerically by taking finite differences. The intercept at $L=0$, ${\tilde G}_{\rm if}$, allows us to extract the damping enhancement \cite{Zwierzycki:prb05} but here we focus on the bulk properties and leave consideration of the material dependence of the interface enhancement for later study. The value of $\alpha$ determined from the slope of ${\tilde G}(L)/(\gamma \mu_s A)$ is $0.0046 \pm 0.0001$ that is at the lower end of the range of values $0.004 - 0.013$ measured at room temperature for Py
\cite{Mizukami:jmmm01,*Mizukami:jjap01,Bailey:ieeem01,
Patton:jap75,*Ingvarsson:apl04,*Nakamura:jjap04,*Rantschler:ieeem05,*Bonin:jap05,
*Lagae:jmmm05,*Nibarger:apl03,*Inaba:ieeem06,*Oogane:jjap06}.

Fig.~\ref{fig:Fig2} shows the Gilbert damping parameter as a function of $x$ for Ni$_{1-x}$Fe$_x$ binary alloys in the fcc structure. From a large value for clean Ni, it decreases rapidly to a minimum at $x \sim 0.65$ and then grows again as the limit of clean {\em fcc} Fe is approached. Part of the decrease in $\alpha$ with increasing $x$ can be explained by the increase in the magnetic moment per atom as we progress from Ni to Fe. The large values of $\alpha$ calculated in the dilute alloy limits can be understood in terms of conductivity-like enhancement at low temperatures \cite{Bhagat:prb74,*Heinrich:jap79} that has been explained in terms of intraband scattering \cite{Kambersky:cjp76,Gilmore:prl07,*Gilmore:jap08,*Kambersky:prb07}. The trend exhibited by the theoretical $\alpha(x)$ is seen to be reflected by experimental room temperature results. In spite of a large spread in measured values, these seem to be systematically larger than the calculated values. Part of this discrepancy can be attributed to an increase in $\alpha$ with temperature \cite{Bastian:pssa76,Bailey:ieeem01}. 

% subsection damping (end)

\begin{figure}[t]
\includegraphics[scale=\scalefig]{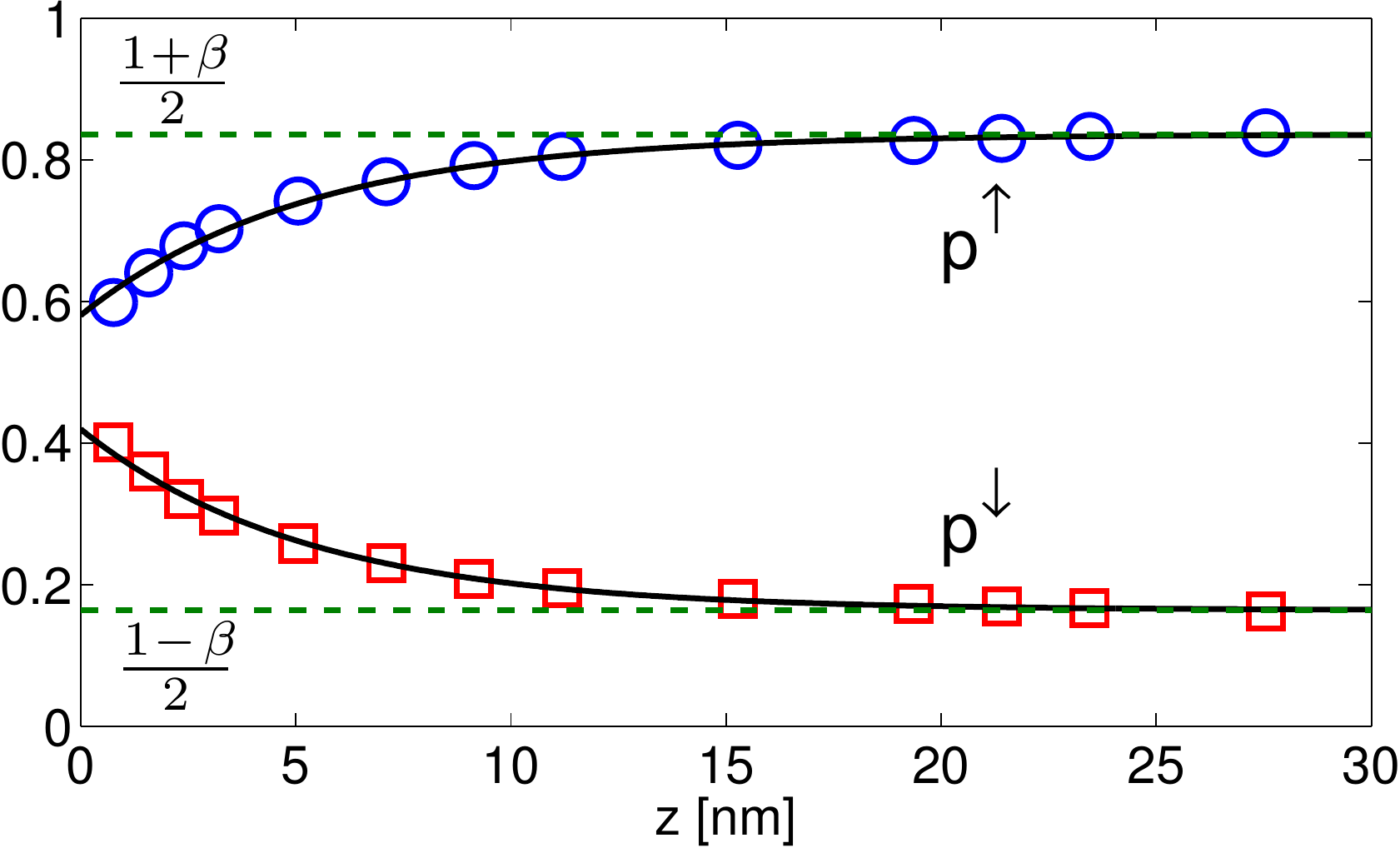}
\caption{Fractional spin-current densities for electrons injected at $z=0$ from Cu into Ni$_{80}$Fe$_{20}$ alloy.
Calculated values (symbols) and fits to Eq.~\eqref{eqn:puda} (solid lines).
}
\label{fig:Ni80Fe20sdif}
\end{figure}

\paragraph{\color{red} Spin diffusion.}When an unpolarized current is injected from a normal metal into a ferromagnet, the polarization will return to the value characteristic of the bulk ferromagnet sufficiently far from the injection point, provided there are processes which allow spins to flip. Following Valet-Fert \cite{Valet:prb93} and assuming there is no spin-flip scattering in the N leads, we can express the fractional spin current densities $p^{\uparrow(\downarrow)}=J^{\uparrow(\downarrow)}/J$ as a function of distance $z$ from the interface as
\begin{equation}
p^{\uparrow(\downarrow)}(z) =\frac{1}{2} \pm \frac{\beta}{2}
\left[1-\frac{exp(-z/l_{\rm sf}) r^{*}_{\rm if} (\beta -\gamma+\gamma \sech{\delta} )} {\beta (r^{*}_{\rm if} + l_{\rm sf} \delta  \rho^{*}_{\rm F} \tanh{\delta})}\right], \label{eqn:puda}
\end{equation}
where $J$ is the total current through the device, $J^{\uparrow}$ and $J^{\downarrow}$ are the currents of majority and minority electrons, respectively, $l_{\rm sf}$ is the spin-diffusion length, $\rho^{*}_{\rm F} = (\rho^{\downarrow} + \rho^{\uparrow})/4$ is the bulk resistivity and $\beta$ is the bulk spin asymmetry  $(\rho^{\downarrow} - \rho^{\uparrow}) / (\rho^{\downarrow} + \rho^{\uparrow})$. The interface resistance $r^{*}_{\rm if}=(r^{\downarrow}_{\rm if} + r^{\uparrow}_{\rm if})/4$, the interface resistance asymmetry $\gamma=(r_{\rm if}{\downarrow} - r_{\rm if}^{\uparrow}) / (r_{\rm if}^{\downarrow} + r_{\rm if}^{\uparrow})$ and the interface spin-relaxation expressed through the spin-flip coefficient $\delta$ \cite{Park:prb00} must be taken into consideration resulting in a finite polarization of current injected into the ferromagnet.
The corresponding expressions are plotted as solid lines in Fig.~\ref{fig:Ni80Fe20sdif}.

To calculate the spin-diffusion length we inject non-polarized states from one N lead and probe the transmission probability into different spin-channels in the other N lead for different thicknesses of the ferromagnet. Fig.~\ref{fig:Ni80Fe20sdif} shows that the calculated values can be fitted using expressions \eqref{eqn:puda} if we assume that $J^{\sigma}/J=G^{\sigma}/G$, yielding values of the spin-flip diffusion length $l_{\rm sf}=5.5\pm0.3$~nm and bulk asymmetry parameter $\beta=0.678 \pm 0.003$ for Ni$_{80}$Fe$_{20}$ alloy compared to experimentally estimated values of $0.7 \pm 0.1$ for $\beta$ and in the range $5.0-6.0$~nm for $l_{\rm sf}$ \cite{Bass:jmmm99,*Bass:jpcm07}. 

\begin{figure}[!t]
\begin{center}
\includegraphics[scale=\scalefig]{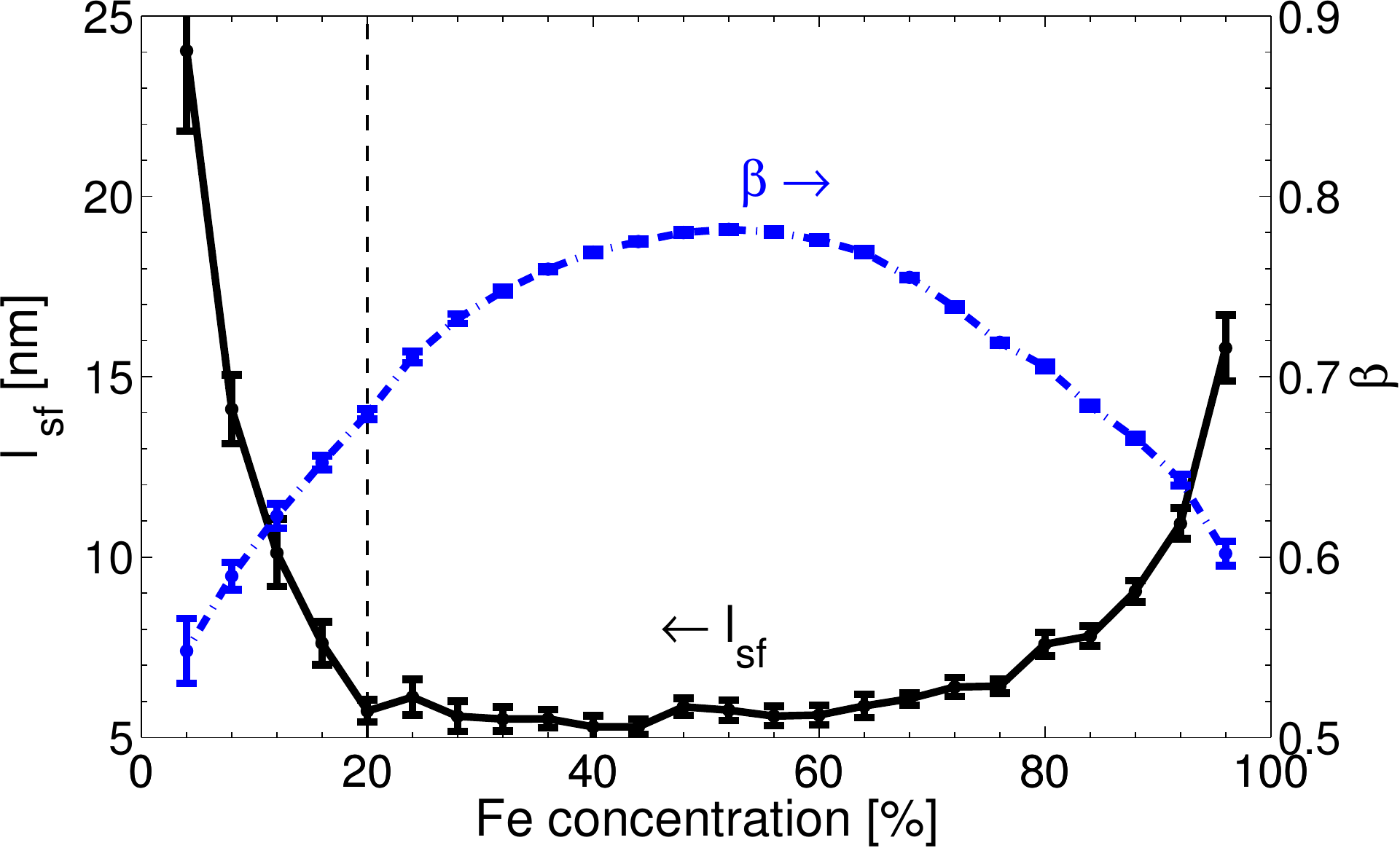}
\end{center}
\caption{Spin-diffusion length (solid line) and polarization $\beta$ as a function of the concentration $x$ for  Ni$_{1-x}$Fe$_x$ binary alloys. } 
\label{fig:NiFe-sdif}
\end{figure}

$l_{\rm sf}$ and $\beta$ are shown as a function of concentration $x$ in Fig.~\ref{fig:NiFe-sdif}. The convex behaviour of $\beta$ is dominated by and tracks the large minority spin resistivity $\rho^{\downarrow}$ whose origin is the large mismatch of the Ni and Fe minority spin band structures that leads to a $\sim x(1-x)$ concentration dependence of $\rho^{\downarrow}(x)$ \cite{Banhart:epl95}. The majority spin band structures match well so $\rho^{\uparrow}$ is much smaller and changes relatively weakly as a function of $x$. The increase of $l_{\rm sf}$ in the clean metal limits corresponds to the increase of the electron momentum and spin-flip scattering times in the limit of weak disorder.

In summary, we have developed a unified DFT-based scattering theoretical approach for calculating transport parameters of concentrated alloys that depend strongly on spin-orbit coupling and disorder and have illustrated it with an application to NiFe alloys. Where comparison with experiment can be made, the agreement is remarkably good offering the prospect of gaining insight into the properties of a host of complex but technologically important magnetic materials.

\begin{acknowledgments}
This work is part of the research programs of 
``Stichting voor Fundamenteel Onderzoek der Materie'' (FOM) 
and the use of supercomputer facilities was sponsored by the 
``Stichting Nationale Computer Faciliteiten'' (NCF), both financially supported by the 
``Nederlandse Organisatie voor Wetenschappelijk Onderzoek'' (NWO).
It was also supported by ``NanoNed'', a nanotechnology programme of
the Dutch Ministry of Economic Affairs and by EC Contract No. IST-033749 “DynaMax.”
\end{acknowledgments}

\providecommand{\noopsort}[1]{}\providecommand{\singleletter}[1]{#1}%


\begin{thebibliography}{50}
\bibitem [{fn2()}]{fn2}%
  \BibitemOpen
  \href@noop {} {}\bibinfo {note} {The non-linearity of the resistance for $L
  \lesssim 20$ in the inset to Fig.~\ref{fig:Fig1} has nothing to do with
  spin-orbit coupling. It already results when spin-dependent resistances of
  the form $ R^{\sigma}(L) = 2/G_{\rm Sh} + 2 R_{\rm if}^{\sigma}+\rho^{\sigma}
  L $ are added in parallel in the two-current series-resistor (2CSR) model;
  the total resistance only becomes linear for values of $L$ so large that the
  bulk resistances $\rho^{\sigma} L$ are much larger than the interface and
  Sharvin terms.}\BibitemShut {Stop}%
\end{thebibliography}
\end{document}